\documentstyle[prl,aps,twoside]{revtex}
\evensidemargin=\oddsidemargin
\begin{document}
\draft
\twocolumn[\hsize\textwidth\columnwidth\hsize\csname@twocolumnfalse%
\endcsname
\preprint{TUIMP-TH-98/91}
\title{Single Top Quark Production in $e\gamma$ Collisions and Testing
 Technicolor Models}
\bigskip
\author{Xuelei Wang}
\address{Department of Physics, Tsinghua University, Beijing 100084, China;\\
Physics Department, Henan Normal University, Xinxiang Henan 453002 
, China.}
\author{Yu-Ping Kuang~~~~ Hong-Yi Zhou}
\address{China Center of Advanced Science and Technology (World 
Laboratory),~
~~~~~~~~~~\\ P.O.Box 8730, Beijing 100080, China;\\
Institute of Modern Physics, Tsinghua University, Beijing
100084, China.\footnote{Mailing address.}}
\author{Hua Wang~~~~ Ling Zhang}
\address{Department of Physics, Tsinghua University, Beijing 100084, China.}
\bigskip\bigskip
\date{TUIMP-TH-98/91}
\maketitle
\begin{abstract}
We study the single top quark production process $e^+ \gamma 
\to t \bar{b}\bar{\nu}_e$ in various kinds of technicolor models in
high energy $e\gamma$ collisions at the future $e^+e^-$ linear colliders. 
It is shown that if there is certain charged pseudo 
Goldstone boson (PGB) coupling to $t\bar{b}$, the $t\bar{b}$-channel PGB 
contribution is dominant, but the situation is quite different
from that in the neutral channel $\gamma\gamma\to t\bar{t}$ due to
the destructive nature of different amplitudes. 
At the DESY linear collider TESLA, the event rates
in models with $t\bar{b}$-channel PGB contributions, such as the 
top-color-assisted technicolor model, etc. are experimentally measurable. 
The $e^+\gamma\to t\bar{b}\bar{\nu}_e$ process provides a feasible test of 
technicolor models with $t\bar{b}$-channel charged PGB contributions.

\end{abstract}
\vspace{0.2cm}\hspace{1.84cm}
PACS number(s): 14.65.Ha, 12.15.Lk, 12.60.Nz
\bigskip\bigskip\bigskip\bigskip\bigskip
]
\begin{center}
{\bf I. INTRODUCTION}
\end{center}
 
So far the most unclear part of the standard model is its symmetry
breaking sector. Probing the electroweak symmetry breaking mechanism
will be one of the most important tasks at future high energy colliders.
Dynamical electroweak symmetry breaking, for example
technicolor type theories, is an attractive idea that it avoids the 
shortcomings of triviality and unnaturalness arising from the
elementary Higgs field. The simplest QCD-like extended technicolor
 model \cite{ETC} leads to a too large oblique correction $S$
parameter \cite{STU}, and is already ruled out by the recent LEP precision 
electroweak measurement data \cite{PDG,KEK}.
Various improvements have been made to make the predictions consistent
with the LEP precision measurement data and even to give possible dynamical 
explanation of the heaviness of the top quark, for example the walking 
technicolor models \cite{WTC},  the Appelquist-Terning one-family model 
\cite{AT}, the multiscale walking technicolor models \cite{MWTC}, the 
top-color-assisted technicolor models 
(TC2) \cite{TC2,Lane},
the non-commuting extended technicolor model 
\cite{NCETC}, etc. 
This kind of dynamical electroweak symmetry breaking theory is one of the 
important candidates of the electroweak symmetry breaking mechanism. Due to 
the strong interaction nature, it is hard to make 
precise calculations in such dynamical electroweak symmetry breaking theories. 
However, there are some characteristic features in this kind of models, for 
instance the prediction for certain pseudo Goldstone bosons (PGB's) in the few 
hundred GeV region. It is thus interesting to find experimentally measurable 
processes which are sensitive to the PGB's to test this kind of models.

The top quark is the heaviest particle yet experimentally discovered
and its mass $175$ GeV \cite{top} is close to the electroweak symmetry 
breaking scale $246$ GeV. Thus processes containing the top quark may be 
sensitive to the electroweak symmetry breaking mechanism, and top quark 
productions at high energy 
colliders can be good processes for probing the electroweak symmetry breaking 
mechanism. There have been many papers studying the test of new physics via top 
quark productions at high energy colliders in the literatures. For instance, 
the model-independent studies in the effective Lagrangian
formalism have been given in Refs.\cite{yuan,young,Boos},
supersymmetric corrections to top quark productions at hadron colliders
and electron (photon) linear colliders (LC) have been studied in Ref.
\cite{SUSYcorr}, top quark pair productions at hadron colliders and photon 
colliders in various technicolor models have been studied in Refs.
\cite{TChadtt,phphtt}.
Recently, there have been a lot of interests in 
studying single top quark productions which provides a sensitive measurement 
of the $Wtb$ coupling \cite{SW}. There have been many papers studying single 
top quark productions in new physics models \cite{1-top}, and also 
model-independent study as well \cite{Boos}. In Refs.\cite{TChadtt,phphtt}
it is shown that, in the $t\bar{t}$ productions, once there is a 
neutral PGB coupling strongly to $t\bar{t}$, the $t\bar{t}$-channel PGB 
contribution is large and dominant over all other loop corrections, and thus 
such processes can be sensitive tests of the neutral PGB effects. In this 
paper, we shall study the $e^+\gamma\to t\bar{b}\bar{\nu}_e$ process at the 
high energy $e\gamma$ colliders in various technicolor models, and we shall 
show that this process is sensitive to the $t\bar{b}$-channel charged PGB 
effects if there is certain charged PGB(s) coupling strongly to
$t\bar{b}$\footnote{The present study is different from that in 
Ref.\cite{Boos} in the following sense. Ref.\cite{Boos} studied 
single top quark production in $e\gamma$ collision at tree level in the
standard model and 
with the consideration of possible model-independent anomalous $Wtb$ couplings 
but without considering the contributions from light PGB(s). Our present
study takes account of one-loop corrections and the light PGB(s) contributions 
in various kinds of technicolor models which are shown to be significant.}.  
We shall see that the $e^+\gamma\to t\bar{b}\bar{\nu}_e$ process at the 
$e\gamma$ colliders based on the LC, especially the DESY TeV Energy 
Superconducting Linear Accelerator (TESLA), is a feasible 
test of the charged PGB effects, and {\it the situation is quite different 
from that in the neutral $\gamma\gamma\to t\bar{t}$ channel due to the large 
mass difference between the top quark and the bottom quark and the
destructive nature of different amplitudes}. Of special 
interest is the test of the charged top-pion effect in the 
TC2 models. The recent Fermilab CDF data on $t\bar{t}$ production at the 
Fermilab Tevatron show that the branching fraction for a top quark decaying 
into a final state $e$ or $\mu$ is consistent with the tree-level
standard model prediction up to a certain uncertainty \cite{CDF}. Future 
improved experiments may lead to more precise conclusion. As has been 
discussed in Ref.\cite{TC2} that this means a charged top-pion lighter than 
the top quark may not be favored, and a charged top-pion heavier than the top 
quark will have a broad width so that it is difficult to detect. Our result 
shows that the $e^+\gamma\to t\bar{b}\bar{\nu}_e$ process at the DESY TESLA 
provides a possible test of the charged top-pion effect. 

This paper is organized as follows. In Sec. II, we shall present the
calculations of the $e^+\gamma\to t\bar{b}\nu_e$ production amplitudes in 
several currently improved technicolor models. We take the TC2 model as
a typical example of models containing charged PGB's strongly coupling to 
$t\bar{b}$. The numerical 
results of the cross sections will be presented in Sec. III, 
and Sec. IV is a concluding remark.
\null\vspace{1.4cm}
\begin{center}
{\bf II. THE $e^+\gamma\to t\bar b\bar \nu_e$ PRODUCTION AMPLITUDES}
\end{center}

The tree-level standard model contributions to the process
\begin{eqnarray}                                                
e^++ \gamma\to t+\bar{b}+\bar{\nu}_e
\end{eqnarray}
are shown in Fig.~1(a)-Fig.~1(d). The momenta of these particles will be 
denoted by $p_{e^+},~p_\gamma,~p_t,~p_{\bar{b}}$ and $p_{\bar{\nu}}$, 
respectively. In high energy processes, the effects of $m_e$
is negligibly small. Thus we neglect it in the following
calculations.
We take the unitary gauge. The obtained tree-level standard model amplitude 
is explicitly
\begin{eqnarray}                                                
{\cal M}_0={\cal M}_0^{(1a)}+{\cal M}_0^{(1b)}+{\cal M}_0^{(1c)}+{\cal M}_0^{(1d)}
\,,
\end{eqnarray}
where
\begin{eqnarray}                                                
&& {\cal M}_0^{(1a)}=-\frac{8i\sqrt{2\pi \alpha}M^2_WG_F }{6}
G(p_{\bar{b}}-p_\gamma;m_b)\\ \nonumber
&&\hspace{1cm}G(p_{e^+}-p_{\bar{\nu}};M_W)
~\bar{u}_t\gamma_{\mu}L(\rlap/p_{\bar{b}}-\rlap/p_\gamma+m_b)
\rlap/\epsilon ~v_b
\hspace{1cm}\\ \nonumber
&&\hspace{1cm}T^W_{\mu\lambda}(p_{e^+}-p_{\bar{\nu}})
~\bar{v}_e\gamma_{\lambda}Lv_{\nu_e}\,,
\end{eqnarray}
\begin{eqnarray}                                                 
&&{\cal M}_0^{(1b)}=-\frac{8i\sqrt{2\pi \alpha}M^2_WG_F}{3}G(p_t-p_\gamma;m_t)
\\ \nonumber
&&\hspace{1cm}G(p_{e^+}-p_{\bar{\nu}};M_W)~\bar{u}_t\epsilon\llap/(p\llap/_t-
p\llap/_\gamma+m_t)\gamma_{\mu}Lv_b\\ \nonumber
&&\hspace{1cm}T^W_{\mu\lambda}(p_{e^+}-p_{\bar{\nu}})
~\bar{v}_e\gamma_{\lambda}Lv_{\nu_e}\,,
\end{eqnarray}
\begin{eqnarray}                                                 
&&{\cal M}_0^{(1c)}=-\frac{8i\sqrt{2\pi \alpha}M^2_WG_F}{2}G(p_{e^+}+p_\gamma;0)
\\ \nonumber
&&\hspace{1cm}G(p_t+p_{\bar{b}};M_W)~\bar{u}_t\gamma_{\rho}Lv_b
~T^W_{\rho \sigma}(p_t+p_{\bar{b}})
\hspace{0.8cm}\\ \nonumber
&&\hspace{1cm}\bar{v}_e\epsilon\llap/
(p\llap/_{e^+}+p\llap/_\gamma)\gamma_{\sigma}Lv_{\nu_e}\,,
\end{eqnarray}
\begin{eqnarray}                                                 
&&{\cal M}_0^{(1d)}=-\frac{8i\sqrt{2\pi \alpha}M^2_WG_F}{2}
G(p_t+p_{\bar{b}};M_W)\\ \nonumber
&&\hspace{1cm}G(p_{e^+}-p_{\bar{\nu}};M_W)~\bar{u}_t\gamma_{\rho}Lv_b
~T^W_{\rho\sigma}(p_t+p_{\bar{b}})\hspace{0.5cm}\\ \nonumber
&&\hspace{1cm}[\epsilon_{\mu} (p_{e^+}-p_{\bar{\nu}}-p_\gamma)_{\sigma}
+\epsilon_{\sigma}(p_t+p_{\bar{b}}+p_\gamma)_{\mu}
\\ \nonumber
&&\hspace{1cm}-g_{\mu\sigma}(p_{e^+}-p_{\bar{\nu}}+p_t+p_{\bar{b}})\cdot
\epsilon]\\ \nonumber
&&\hspace{1cm}T^W_{\mu\lambda}(p_{e^+}-p_{\bar{\nu}})
~\bar{v}_e\gamma_{\lambda}Lv_{\nu_e}\,,
\end{eqnarray}
with the propagator
\begin{eqnarray}                                                   
G(p;M)\equiv \frac{1}{p^2-M^2}\,,
\end{eqnarray}
the tensor
\begin{eqnarray}                                                    
T^W_{\rho\sigma}(p_i+p_j)\equiv g_{\rho\sigma}-\frac{(p_i+p_j)_{\rho} 
(p_i+p_j)_{\sigma}}{M^2_W}\,,
\end{eqnarray}
and $L\equiv \frac{1}{2}(1-\gamma_5),~R\equiv \frac{1}{2}(1+\gamma_5)$. In
(7), $M$ stands for the mass of the particle.

The technicolor and top-color contributions to this process 
depends on the models. We take certain models as typical examples.\\

\null\noindent
{\bf 1. The TC2 Models}\\

We first consider the TC2 model. There have been improvements of the TC2 model
\cite{Lane} to overcome some shortcomings of the original model
and make it more realistic \cite{short,Lane}.
Since the purpose of this paper is to test the characteristic effects of the 
charged PGB's, we are not considering the delicate refinements and 
shall simply take the original TC2 model (it will be refered as model
TC2-I in this paper) \cite{TC2} and the
topcolor-assisted multiscale technicolor model (it will be refered as
model TC2-II in this paper) model
\cite{TOPCMTC,TChadtt,phphtt}
as typical examples.\\

In model TC2-I, there is a charged top-pion $\Pi_t^+$ in the
top-color sector with mass roughly around $200$ GeV and decay constant
$F_{\Pi_t}=50$ GeV \cite{TC2}. We take its technicolor sector to be the 
standard extended technicolor model, thus there is a charged technipion 
$\Pi^+$ from the technicolor sector with mass roughly around $100$ GeV (or
larger) and decay constant $F_\Pi=123$ GeV \cite{ETC}. Model TC2-II 
\cite{TOPCMTC,TChadtt,phphtt}
differs from model TC2-I only by 
its extended technicolor sector which is taken to be the multiscale walking 
technicolor model \cite{MWTC} in which the technipion decay constant is 
$F_\Pi=40$ GeV \cite{MWTC} rather than $123$ GeV. To see the influence of the 
PGB masses on the production cross section, we take $m_{\Pi_t}$ and $m_{\Pi}$ 
to vary in certain ranges: $150~{\rm GeV}\le m_{\Pi_t} \le 380~{\rm GeV}$ and
$100~{\rm GeV} \le m_{\Pi} \le 220~{\rm GeV}$. The values of these
parameters are summarized in Table 1.

\null\noindent
{\bf ~~a. Model TC2-I}\\

As we have seen from Refs.\cite{tdecay,phphtt} that
the technicolor and top-color gauge boson contributions to the $Wtb$ vertex and
the direct technicolor dynamics contribution to the $t\bar{t}$ production rate 
are only of the order of a few percent which are much smaller than the
resonance enhanced $t\bar{t}$-channel PGB contributions to the top quark pair 
productions \cite{TChadtt,phphtt}.
Thus we concentrate our study to the 
PGB contributions in this paper. 

The Feynman diagrams for PGB contributions to the process $e^+\gamma\to
t\bar{b}\bar{\nu}_e$ are shown in Fig.~2-Fig.~3\footnote{There are 
additional loop diagrams with the photon line attached to the PGB line and the
external $t$ and $\bar{b}$ lines which vanish
in the approximation $m_e\approx 0$ in our calculation,
so that they are not shown in Fig.~1-Fig.~3.}. Fig.~2, enhanced by the 
PGB resonance effects, are the
most important $t\bar{b}$-channel PGB contributions.
The contributions from Fig~3, without PGB resonance enhancements, 
are only of the order of ordinary radiative corrections (less than $1\%$) and 
is negligibly small compared with those from Fig.~2.

The technifermion triangle loop contribution to the $\Pi^+-W^+-\gamma$ vertex
[Fig.~2(a)] in extended technicolor models can be approximately evaluated 
\cite{Lubicz} from 
the formulae for the Adler-Bell-Jackiw anomaly \cite{anomaly}. The result has 
been given in Ref. \cite{Ellis}. The $\Pi^+-W^+-\gamma$ coupling is:
\begin{eqnarray}                                                    
\frac{S_{\Pi^+W^+\gamma}}{4\pi^2F_{\pi}}\varepsilon_{\mu\nu\alpha\beta}
\epsilon_1^{\mu}\epsilon_2^{\nu}k_1^{\alpha}k_2^{\beta}\,,
\end{eqnarray}
\begin{eqnarray}                                                    
S_{\Pi^+W^+\gamma}=\frac{e^2}{2\sqrt{3}s_w}N_{TC}\,,
\end{eqnarray}
where the technicolor number is taken to be $N_{TC}=4$.  

The quark triangle loops [Fig.~2(b)-Fig.~2(c)] are more complicated. They 
contain both the top quark and the bottom quark propagators with the masses
of these quarks relatively light and significantly different.
The contributions of these triangle loops are thus essentially different
from the result of the Adler-Bell-Jackiw anomaly, and they actually contain
logarithmic ultraviolet divergences. There are no corresponding tree-level
terms to absorb these divergences. However, to guarantee $U(1)_{em}$
gauge invariance, we should also take account of Fig.~2(d)-Fig.~2(e) 
[nonvanishing as $m_e\approx 0$] which also contain logarithmic ultraviolet 
divergences, and cannot be absorbed into tree-level terms. Explicit results 
show that these two kinds of ultraviolet divergences just cancel each other 
and the total result is finite as it should be. Because of the destructive 
nature of these two kinds of amplitudes, there is a significant cancellation 
between the finite corrections from these two kinds of amplitudes.
This makes the charged channel $e^+\gamma\to t\bar{b}\bar{\nu}_e$ very 
different from the neutral channel $\gamma\gamma\to t\bar{t}$, and the
detection of the charged channel needs a larger integrated luminosity
as we shall see in Sec. III.

To explicitly calculate the contributions of Fig.~2(b)-Fig.~2(e), we need the 
couplings of the technipion $\Pi^+$ and the top-pion $\Pi^+_t$ to quarks. In 
the original extended technicolor models without top-color, the coupling of 
$\Pi^+$ to quarks has been given in Ref.\cite{Ellis} which is 
\begin{eqnarray}                                                    
ic_f\frac{m_t}{F_{\Pi}}~\bar{u}_tLu_b~\Pi^{+}+h.c.\,,
\end{eqnarray} 
where $c_f=\frac{1}{\sqrt{6}}$.
In TC2 models, the extended technicolor dynamics only provides a small part of 
the top quark mass $m'_t$. For reasonable range of the parameters in TC2 
models, $m'_t\sim 5-20$ GeV \cite{TC2,Balaji}.
Thus the coupling of the 
technipion $\Pi^+$ to quarks in TC2 models can be obtained
by replaceing $m_t$ by $m'_t$ in (11), i.e.
\begin{eqnarray}                                                    
ic_f\frac{m'_t}{F_{\Pi}}~\bar{u}_tLu_b~\Pi^{+}+h.c.\,\,.
\end{eqnarray}              
Similarly, the coupling of the top-pion $\Pi^+_t$ to quarks is of the 
following form  
\begin{eqnarray}                                                   
i\frac{m_t-m'_t}{F_{\Pi_t}}~\bar{u}_tLu_b~\Pi^+_t+h.c.\,\,.
\end{eqnarray}

In the $\Pi^+(\Pi^+_t)$-propagator in Fig.~2(a)-Fig.~2(e), the
time-like momentum may hit the $\Pi^+(\Pi^+_t)$-pole. So we should take
into account the effects of the widths of $\Pi^+$ and $\Pi^+_t$ in the 
calculation. For the $\Pi^+(\Pi^+_t)$-propagator in (7), we take the 
complex mass term $M^2-iM\Gamma$ instead of the simple mass term $M^2$
to include the effect of the width $\Gamma$ of $\Pi^+(\Pi^+_t)$.
The $-iM\Gamma$ term is important in the vicinity of the resonance. It
is this resonance contribution that enhances the amplitudes of 
Fig.~2(a)-Fig.~2(e). When $M_{\Pi},~ M_{\Pi_t}>m_t$,~
the dominant decay mode of $\Pi^+$ and $\Pi^+_t$ is $t\bar{b}$.
So, in this case the widths $\Gamma_{\Pi}$ and $\Gamma_{\Pi_t}$ are
\begin{eqnarray}                                                    
\Gamma_{\Pi}\approx \Gamma_{\Pi^+}(\Pi \to t\bar{b})
=c_f^2\frac{m_t^{'2}(m^2_{\Pi}-m_t^2)^2}{16\pi F^2_{\Pi}
M^3_{\Pi}}
\end{eqnarray}  
and
\begin{eqnarray}                                                    
\Gamma_{\Pi_t}\approx \Gamma_{\Pi_t^+}(\Pi_t\to t\bar{b})
=\frac{(m_t-m'_t)^2(m^2_{\Pi_t}-m_t^2)^2}{16\pi F^2_{\Pi_t}
M^3_{\Pi_t}} \,.
\end{eqnarray}  
When $M_{\Pi_t},~M_{\Pi}<m_t$,
$\Pi_t^+$ and $\Pi^+$ decay dominantly into $c\bar{s}$.
For small $m_b,~m_s$, we can approximately take $\Gamma_{\Pi_t}^{}$ and
$\Gamma_{\Pi}^{}$ to be zero.

With (12)-(15), we can do the explicit calculation of the
contributions of Fig.~2(b)-Fig.~2(e) to the amplitude.
In the calculation, we take dimensional regularization and the on-shell
renormalization scheme. The obtained amplitude with technicolor corrections is
\begin{eqnarray}                                                   
&&{\cal M}={\cal M}_0+\Delta {\cal M}_{TC}^{(2a)}
+\Delta {\cal M}_{TC}^{(2b-2e)}(\Pi^+)\\ \nonumber
&&\hspace{0.8cm}+\Delta {\cal M}_{TC}^{(2b-2e)}(\Pi_t^+)
+\Delta {\cal M}_{TC}^{(3a-3i)}\,,
\end{eqnarray}
where the superscripts denote the corresponding Feynman diagrams in Fig.~2.
The explicit formulae for $~\Delta {\cal M}_{TC}^{(2a)},
~\Delta{\cal M}_{TC}^{(2b-2e)}(\Pi^+),~{\rm and}
~\Delta{\cal M}_{TC}^{(2b-2e)}(\Pi_t^+)~$ are
\begin{eqnarray}                                                    
&&\Delta{\cal M}_{TC}^{(2a)}=-c_f\frac{m'_tM_WS_{\Pi^+W^+\gamma}}
{4\pi^2F_{\Pi}^2}\sqrt{2\sqrt{2}G_F}\\ \nonumber
&&\hspace{1cm}G(p_t+p_{\bar{b}};M_\pi)G(p_{e^+}-p_{\bar{\nu}};
M_W)\\ \nonumber
&&\hspace{1cm}\bar{u}_tLv_b~
\varepsilon^{\mu\nu\alpha\beta}
\epsilon_{\mu}(p_{\gamma)\alpha}(p_{e^+}-p_{\nu})_{\beta}\\ \nonumber
&&\hspace{1cm}T^W_{\mu\lambda}(p_{e^+}-p_{\bar{\nu}})~\bar{v}_{e}
\gamma_{\lambda}Lv_{\nu_e}\,,
\end{eqnarray}
\begin{eqnarray}                                                    
&&\Delta{\cal M}_{TC}^{(2b-2e)}(\Pi
^+)=-ic_f\frac{m'_tM_W}{F_{\Pi}}
\sqrt{2\sqrt{2}G_F}\\ \nonumber
&&\hspace{1cm}G(p_t+p_{\bar{b}};M_{\Pi})
G(p_{e^+}-p_{\bar{\nu}};M_W)~\bar{u}_tLv_b\hspace{1.5cm}\\ \nonumber
&&\hspace{1cm}\left[\Gamma^{(2b)}_{\mu\nu}(\Pi)
+\Gamma^{(2c)}_{\mu\nu}(\Pi)+\Gamma^{(2d)}_{\mu\nu}(\Pi)\right]\\ \nonumber
&&\hspace{1cm}\epsilon_{\nu}
T^W_{\mu\lambda}(p_{e^+}-p_{\bar{\nu}})
~\bar{v}_{e}\gamma_{\lambda}Lv_{\nu_e}\\ \nonumber
&&\hspace{1cm}
+ic_f\frac{m'_tM_W}{F_{\Pi}}\sqrt{8\pi\sqrt{2}G_F\alpha}\\ \nonumber
&&\hspace{1cm}G(p_t+p_{\bar{b}};M_{\Pi})G(p_t+p_{\bar{b}};M_W)
G(p_{e^+}+p_\gamma;0)\\ \nonumber
&&\hspace{1cm}\bar{u}_tLv_b~
\left[-i\Sigma_{\mu}(\Pi)\right]~T^W_{\mu\nu}(p_t+p_{\bar{b}})\\ \nonumber
&&\hspace{1cm}\bar{v}_{e}\rlap/\epsilon(\rlap/p_{e^+}+\rlap/p_\gamma)
\gamma_{\nu}Lv_{\nu_e}\,,\nonumber
\end{eqnarray}
and  
\begin{eqnarray}                                                    
&&\Delta{\cal M}_{TC}^{(2b-2e)}(\Pi_t^+)=-i\frac{(m_t-m'_t)M_W}{F_{\Pi_t}}
\sqrt{2\sqrt{2}G_F}\\ \nonumber
&&\hspace{1cm}G(p_t+p_{\bar{b}};M_{\Pi_t})G(p_{e^+}-p_{\bar{\nu}};M_W)
~\bar{u}_tLv_b\\ \nonumber
&&\hspace{1cm}\left[\Gamma^{(2b)}_{\mu\nu}(\Pi_t)
+\Gamma^{(2c)}_{\mu\nu}(\Pi_t)+\Gamma^{(2d)}_{\mu\nu}(\Pi_t)\right]
\epsilon_{\nu}\\ \nonumber
&&\hspace{1cm}T^W_{\mu\lambda}(p_{e^+}-p_{\bar{\nu}})~\bar{v}_{e}\gamma_{\lambda}
Lv_{\nu_e}\\ \nonumber
&&\hspace{0.5cm}+i\frac{(m_t-m'_t)M_W}{F_{\Pi_t}}
\sqrt{8\pi\sqrt{2}G_F\alpha}~G(p_t+p_{\bar{b}};M_{\Pi_t})\\ \nonumber
&&\hspace{1cm}G(p_t+p_{\bar{b}};M_W)G(p_{e^+}
+p_\gamma;0)
~\bar{u}_tLv_b\\ \nonumber
&&\hspace{1cm}\left[-i\Sigma_{\mu}(\Pi_t)\right]~T^W_{\mu\nu}(p_t+p_{\bar{b}})
~\bar{v}_{e}\rlap/\epsilon(\rlap/p_{e^+}+\rlap/p_\gamma)\gamma_{\nu}Lv_{\nu_e}
\,.
\nonumber
\end{eqnarray}
In (18)-(19) the functions $\Gamma^{(2b)}_{\mu\nu},~\Gamma^{(2c)}_{\mu\nu},
~\Gamma^{(2d)}_{\mu\nu},~{\rm and}~\Sigma_\mu$ are given in the
Appendix in terms of the standard 2-point and 3-point functions
$B_0,~B_1$ and $~C_0,~C_{ij}$ of the Feynman integrals \cite{PV}. The 
formula for $\Delta {\cal M}_{TC}^{(3a-3i)}$ is quite lengthy and we are not
going to show it since its contribution is only of the order of
ordinary radiative corrections (less than $1\%$) and is negligibly small 
compared with those shown in eqs.(17)-(19) which are enhanced by the PGB 
resonance effects.\\

\null\noindent
{\bf ~~b. Model TC2-II}\\

 In model TC2-II, the extended technicolor sector is taken to be the multiscale
walking technicolor model \cite{MWTC} in which the technipion $\Pi^+$ is 
almost composed of pure techniquarks \cite{TOPCMTC}, 
Thus the relevant changes in the above formulae are: 
\begin{eqnarray}                                                       
c_f=\frac{2}{\sqrt{6}},~~~~S_{\Pi^+W^+\gamma}=\frac{e^2}{4\sqrt{3}s_w}\,.
\end{eqnarray}
The smallness of the decay constant $F_{\Pi}$ in this model [cf. Table 1]
will enhance the $\Pi^+$ contribution [cf. eqs.(17)-(19)].\\

\null\noindent
{\bf 2. The Appelquist-Terning One Family\\ \null\hspace{0.4cm}
 Extended Technicolor Model}\\

This model is designed in which the techniquark sector respects the
custodial $SU(2)$ symmetry, while the technilepton sector is custodial
$SU(2)$ violating. The vacuum expectation value (VEV) $~F_Q~$ of the 
techniquark condensate is much larger than the VEV $~F_L~$ of the technilepton
condensate \cite{AT}. There are $~36~$ PGB's in this model, and the color 
singlet PGB's are mainly composed of technileptons which is irrelevant to the
production of $t\bar{b}$. Thus in this model there are no $~\Delta {\cal
M}_{TC}^{(2a)}(\Pi^+)~$ and $~\Delta{\cal M}_{TC}^{(2b-2e)}(\Pi^+)~$. The only
technicolor contribution to the $e^+\gamma\to t\bar{b}\bar{\nu}_e$ is
$~\Delta{\cal M}_{TC}^{(3a-3i)}~$ which is much smaller than those in (17)-(19).
Thus the cross sections in this model will be much smaller than those
in the previous models.\\

We shall see from the numerical results in the next section that, for certain
parameter range, these models can all be measured and distinguished by their 
$e^+\gamma\to t\bar{b}\bar{\nu}_e$ rates at the DESY TESLA.

\null\noindent
\vspace{0.4cm}
\begin{center}
{\bf III. THE CROSS SECTIONS}
\end{center}

The hard photon beam of the $e^+\gamma$ collider can be obtained from
laser backscattering at the $e^+e^-$ linear collider \cite{photon}. Let
$\hat{s}$ and $s$ be the center-of-mass energies of the $e^+\gamma$ and
$e^+e^-$ systems, respectively. After calculating the cross section 
$\sigma(\hat{s})$ for the subprocess $e^+\gamma\to t\bar{b}\bar{\nu}_e$,
the total cross section at the $e^+e^-$ linear collider can be obtained
by folding $\sigma(\hat{s})$ with the photon distribution function 
$f_\gamma(x)$ ($\hat{s}=xs$)
\begin{eqnarray}                                                   
\sigma_{tot}=\int\limits_{(m_t+m_b)^2/s}^{x_{max}}dx\hat{\sigma}
(\hat{s})f_{\gamma}(x)\,,
\end{eqnarray}
where
\begin{eqnarray}                                                    
\displaystyle
&&f_\gamma(x)=\frac{1}{D(\xi)}\left[1-x+\frac{1}{1-x}
-\frac{4x}{\xi(1-x)}\right.\\ \nonumber
&&\hspace{0.4cm}\left.+\frac{4x^2}{\xi^2(1-x)^2}\right],
\end{eqnarray}
with \cite{photon}
\begin{eqnarray}                                                    
\displaystyle
&&D(\xi)=\left(1-\frac{4}{\xi}-\frac{8}{\xi^2}\right)
\ln(1+\xi)+\frac{1}{2}+\frac{8}{\xi}\\ \nonumber
&&\hspace{1.2cm}-\frac{1}{2(1+\xi)^2}\,.
\end{eqnarray}
In (22) and (23), $\xi=4E_e\omega_0/m_e^2$ in which $m_e$ and $E_e$
stand, respectively, for the incident electron mass and energy,
$\omega_0$ stands for the laser photon energy, and $x=\omega/E_e$ stands
for the fraction of energy of the incident electron carried by the
back-scattered photon. $f_\gamma$ vanishes for $x>x_{max}
=\omega_{max}/E_e=\xi/(1+\xi)$. In order to avoid the creation of
$e^+e^-$ pairs by the interaction of the incident and back-scattered photons,
we require $\omega_0x_{max}\leq m_e^2/E_e$ which implies that $\xi\leq 2+
2\sqrt{2}\approx 4.8$. For the choice of $\xi=4.8$, we obtain
\begin{eqnarray}                                                     
x_{max}\approx 0.83,\hspace{1cm} D(\xi)\approx 1.8 \,.
\end{eqnarray}

  In the calculation of $\sigma(\hat{s})$, instead of calculating the square 
of the renormalized amplitude $~{\cal M}~$ analytically, we calculate the 
amplitudes numerically by using the method of Ref.\cite{HZ}. This greatly 
simplifies our calculations.  Care must be taken in the calculation of the 
form factors expressed in terms of the standard loop integrals defined in 
Ref.\cite{PV}. 
In the numerical calculation, we used the formulae for
the tensor loop integrals given in Ref.\cite{PV}  in which the 
stability of the numerical calculation is poor when the scattering is 
forwards or backwards\cite{Denner}.   
This problem can be avoided by taking certain kinematic cuts on the
rapidity $~y~$ and the transverse momentum $~p_T~$ of the final states 
which are also needed in experimental detections. In order to compare 
with the corresponding
results in the neutral channel, we take, in this paper, the same
kinematic cuts as in Ref.\cite{phphtt}, i.e.
\begin{eqnarray}                                                      
|y|< 2.5,~~~~~~~~~~p_T> 20~\rm{GeV}\,.
\end{eqnarray}
The cuts will also increase the relative correction \cite{Beenakker}.

 In our calculation, we take $m_t=176$ GeV, $m_b=4.9$ GeV, $M_W=80.33$ GeV,
 $G_F=1.19347\times 10^{-5}~{\rm (GeV)}^{-2}$, $s^2_w=0.23$. The 
electromagnetic fine structure constant $\alpha$ at certain energy scale is 
calculated from the simple QED one-loop evolution formula with the boundary 
value $\alpha=1/137.04$ \cite{Donoghue}, and we will not consider the hadronic 
corrections here since it does not affect the conclusions in this paper to the
present precision.

For estimating the event rates, we take the following integrated luminosities
corresponding to a one-year-run at the DESY TESLA 
\cite{Zerwas} 
\begin{eqnarray}                                                       
&&~~\sqrt{s}=0.5~{\rm TeV}:~~~~\int {\cal L}dt\approx 500~{\rm fb}^{-1}\\
\nonumber
&&~~\sqrt{s}=0.8~{\rm TeV}:~~~~\int {\cal L}dt=500~{\rm fb}^{-1}\\
\nonumber
&&~~\sqrt{s}=1.6~{\rm TeV}:~~~~\int {\cal L}dt\geq 500~{\rm fb}^{-1}\,.
\end{eqnarray} 
Our numerical results show that the contributions from the diagrams in
Fig. 3 to the production cross section are negligibly small in all 
models considered in this paper. Therefore we simply ignore them.

In model TC2-I, the numerical results show that the technicolor PGB 
contributions to the production cross section are negligibly small 
compared with the contributions from the top-pion for $m_{\Pi}\alt
220$ GeV. In Table 2, we list the correction to the production cross section 
$\Delta\sigma_{\Pi_t}$ (from the top-pion contributions) and the total cross 
section $\sigma$ with $~150~{\rm GeV}\le m_{\Pi_t}\le 380~{\rm GeV}~$, 
$~m'_t=5,~20~{\rm GeV}~$ at the $0.5$ and $1.6$ TeV LC. We see that the 
correction with $~m_{\Pi_t}=150$ GeV is significantly smaller than those with 
larger $~m_{\Pi_t}~$. This is due to the $~t\bar{b}~$ threshold effect in the 
$~\Pi_t~$ resonance contribution. For $\sqrt{s}=0.5$ TeV, we see from
Table 2 that the relative correction $~\Delta\sigma_{\Pi_t}/\sigma_0~$ is 
around $0.1\%$ if $~m_{\Pi_t}~$ is smaller than the threshold and around 
$(2\sim 4)\%$ if $~m_{\Pi_t}~$ is larger than the threshold. For 
$\sqrt{s}=1.6$ TeV, $~\Delta\sigma_{\Pi_t}/\sigma~$ is around 
$(0.3\sim 0.8)\%$ 
These technicolor
PGB corrections are quite small compared with those in the neutral channel
$\gamma\gamma\to t\bar{t}$ \cite{phphtt}. This is because that the 
contributions of Fig.~2(b)-Fig.~2(c) and Fig.~2(d)-Fig.~2(e) are destructive, 
which makes the charged channel very different from the neutral channel.
With the integrated luminosity in (26), we see from the values of $\sigma$ 
in Table 2 that, for a four-year run, there can be about $\approx 7000$ events 
for $\sqrt{s}=0.5$ TeV, and 
$\geq 30000$ events for $\sqrt{s}=1.6$ TeV. The corresponding statistical 
uncertainties at the $95\%$ C.L. are then $2\%$ for $\sqrt{s}=0.5$ TeV, 
and $\alt 1\%$ for $\sqrt{s}=1.6$ TeV. Thus the effect of the top-pion 
corrections can hardly be experimentally detected if $~m_{\Pi_t}~$ is smaller 
than the threshold and for $\sqrt{s}=1.6$ TeV, but {\it there is a possibility
of detecting the signal for $\sqrt{s}=0.5$ TeV if $~m_{\Pi_t}~$ is larger 
than the threshold} in the sense of statistical uncertainty
\footnote{Since the ordinary one-loop radiative corrections, such as
the contributions from Fig.~3(a)-Fig.~3(i), are already less than $1\%$, the
theoretical uncertainty of this calculation (higher loop effects) is
expected to be unimportant relative to the statistical uncertainty. A 
practical analysis of the detectability concerns also the 
systematic error and the detection efficiency in the experiments, but this is 
beyond the scope of this paper.}. To show the above results more
intuitively, we plot, in Fig.~4 and Fig.~5, the number of events in a
four-year run of the DESY TESLA at 0.5 TeV and 1.6 TeV, respectively,
with $m_{\Pi_t}$ in the range 180 GeV $\alt m_{\Pi_t}\alt$ 380 GeV. In
the figures, the solid line stands for the total number of events, the
dashed line stands for the number of events corresponding to the
tree-level standard model contribution, and the dotted lines indicate the 
bounds of the statistical uncertainty at the $95\%$ C.L. We see that
the signal of $\Pi_t$ contribution can possibly be detected in the range 
200 GeV $\alt m_{\Pi_t}\alt$ 320 GeV at $\sqrt{s}=$ 0.5 TeV, while it cannot 
be detected at the $95\%$ C.L. at $\sqrt{s}=$ 1.6 TeV.

In model TC2-II, the top-pion contributions are similar, while the
$\Pi^+$ contributions are more significant than that in model TC2-I due to the 
smallness of $F_\Pi$ in model TC2-II. The numerical results in model TC2-II 
are listed in Table 3 with the same ranges of $~m_{\Pi_t}~$ and $~m'_t~$, and 
with $~m_{\Pi}= 100~{\rm and}~220$ GeV at the $\sqrt{s}=0.5$ TeV and 
$\sqrt{s}=1.6$ TeV LC. We see that the corrections are also significantly 
different for $~m_{\Pi_t}~$ lying below or above the threshold. 
We see from Table 3 that the effect of $\Pi^+$ contributions, 
$\Delta\sigma_{\Pi}$, is negligibly small for $m_{\Pi}=100$ GeV,
while is almost dominant for $m_{\Pi}=220$ GeV due to the effect of
the tail of the $\Pi^+$-resonance. 
Take the $~m'_t=5$ GeV case as an example. For $~m_{\Pi}=220$ GeV, the 
relative correction $~\Delta\sigma/\sigma~$ ($\Delta\sigma\equiv 
\Delta\sigma_{\Pi_t}+\Delta\sigma_{\Pi}$) is around $16\%$ for 
$\sqrt{s}=0.5$ TeV and is around $10\%$ for $\sqrt{s}=1.6$ TeV, both for 
$~m_{\Pi_t}~$ smaller or larger than the threshold, i.e. the corrections are 
much larger in this model due to the $\Pi^+$ contributions. For 
$~m_{\Pi}=220$ GeV, the relative difference between the cross sections in 
model TC2-II and model TC2-I $~[\sigma({\rm TC2-II})-\sigma({\rm TC2-I})]/
\sigma({\rm TC2-I})~$ is about 
$15\%$ when $\sqrt{s}=0.5$ TeV and about $12\%$ when $\sqrt{s}=1.6$ TeV. In 
this example, the number of events in a four-year run are about $7800$ 
for $\sqrt{s}=0.5$ TeV, and about $33000$ for $\sqrt{s}=1.6$ TeV. 
The corresponding statistical uncertainties at the $95\%$ C.L. are then
$2\%$ for $\sqrt{s}=0.5$ TeV and $1\%$ for $\sqrt{s}=1.6$ TeV. Thus
the effect of the technicolor corrections in model TC2-II can be
clearly detected both at the $\sqrt{s}=0.5$ TeV and $\sqrt{s}=1.6$ TeV 
energies. The difference between models TC2-I and TC2-II
can also be clearly detected at $\sqrt{s}=0.5,~1.6$TeV. So we conclude that 
{\it models {\rm TC2-I} and {\rm TC2-II} can even be experimentally 
distinguished at the $\sqrt{s}=0.5,~1.6$ {\rm TeV TESLA} via 
$e^+\gamma\to t\bar{b}\nu_e$ if $~m_{\Pi}~$ is around $220$ {\rm GeV}}.
The plots corresponding to Fig. 4 and Fig. 5 for model TC2-II are given
in Fig. 6 and Fig. 7, respectively.

As has been mentioned in Sec. II-2, the special arrangement of $F_Q$ and $F_L$
in the Appelquist-Terning one-family walking technicolor model causes that the 
color-singlet technipions are mainly composed of the technileptons, so that 
they do not couple to $t\bar{b}$. Thus there is no $t\bar{b}$-channel PGB 
contribution to the production cross section, and the technicolor corrections 
are only from the diagrams in Fig. 3 which are negligibly small. 
Numerical calculation shows that the relative correction is smaller than $1\%$.
So the effect of the technicolor corrections cannot be detected via the process
$e^+\gamma\to t\bar{b}\nu_e$. This is significantly different from the above
two top-color-assisted technicolor models.

\null\vspace{0.4cm}
\begin{center}
{\bf IV. CONCLUSIONS}
\end{center}

In this paper, we have studied the possibility of testing different
currently interesting improved technicolor models in the process 
$e^+\gamma\to t\bar{b}\bar{\nu}_e$ at the $\sqrt{s}=0.5$~TeV and 
$\sqrt{s}=1.6$~TeV LC, especially the DESY TESLA, via the effects of their 
typical PGB's. We see that the $t\bar b$-channel PGB contributions play 
dominant roles in this production process and their effects are experimentally 
detectable for certain reasonable parameter range in the sense of the 
statistical uncertainty. However, due to the destructive nature of the 
contributions of Fig.~2(b)-Fig.~2(c) and Fig.~2(d)-Fig.~2(e), the relative 
corrections in this charged channel are much smaller than those in the neutral 
channel $\gamma\gamma\to t\bar{t}$ \cite{phphtt}, so that larger integrated
luminosity is needed in the detection.

Specifically, in a four-year run of the DESY TESLA, the effects of the
$t\bar{b}$-channel PGB's in models TC2-I and TC2-II are all
experimentally detectable for reasonable parameter range, and these 
models can be experimentally distinguished through the differences of
their cross sections. The Appelquist-Terning model, as a typical example
of models without a $t\bar{b}$-channel PGB, is not detectable in the
$e^+\gamma\to t\bar{b}\bar{\nu}_e$ process at the LC. 
Thus the $e^+\gamma\to t\bar{b}\bar{\nu}_e$ process at the LC provides
a feasible test of the $t\bar{b}$-channel charged PGB's in various technicolor
models.

Since the recent Fermilab CDF data on $t\bar{t}$ production at the 
Fermilab Tevatron show that the branching fraction for a top quark decaying 
into a final state $e$ or $\mu$ is consistent with the standard model 
prediction up to certain uncertainty \cite{CDF}, a charged top-pion lighter 
than the top quark may not be favored \cite{TC2}, and a charged top-pion 
heavier than the top quark will have a braod width so that it is difficult to 
detect directly. Our results in model TC2-I show that the 
$e^+\gamma\to t\bar{b}\bar{\nu}_e$ process at the DESY TESLA provides feasible 
tests of the charged top-pion effect.

\section*{ACKNOLEDGMENTS}
This work is supported by the National Natural Science
Foundation of China, the Fundamental Research Foundation of Tsinghua 
Univeristy and a special grant from the State Commission of Education
of China.

\null\vspace{0.5cm}
\section*{APPENDIX}

Here we give the explicit expressions for
$\Gamma^{(2b)}_{\mu\nu}(\Pi)$, $\Gamma^{(2c)}_{\mu\nu}(\Pi)$,
$\Gamma^{(2d)}_{\mu\nu}(\Pi)$, $\Sigma_{\rho}(\Pi)$,
$\Gamma^{(2b)}_{\mu\nu}(\Pi_t)$, $\Gamma^{(2c)}_{\mu\nu}(\Pi_t)$,
$\Gamma^{(2d)}_{\mu\nu}(\Pi_t)$, and $\Sigma_{\rho}(\Pi_t)$ which
can be obtained by direct calculations of the 
Feynman diagrams in Figs.~2(c)-2(e). The explicit expressions are
$$                                                                  
\Gamma^{(2b)}_{\mu\nu}(\Pi)=-c_f\frac{M_Wm_tm'_t}
{12\pi^2F_{\Pi}}\sqrt{2\sqrt{2}\pi
G_F\alpha}\bigl\{ 2[(p_{e^+}-p_{\bar{\nu_e}})_{\mu}
$$
$$
\hspace{0.2cm}(p_{e^+}-p_{\bar{\nu_e}})_{\nu}C_{21}+p_{\gamma\mu}p_{\gamma\nu}C_{22}
+(p_{e^+}-p_{\bar{\nu_e}})_{\mu}p_{\gamma\nu}C_{23}
$$
$$
\hspace{0.2cm}+p_{\gamma\mu}(p_{e^+}-p_{\bar{\nu_e}})_{\nu}C_{23}+g_{\mu\nu}C_{24}]
-g_{\mu\nu}B_0(p_{\gamma},m_b,m_b)
$$
$$
\hspace{0.2cm}-g_{\mu\nu}m^2_tC_0+(2p_{e^+}-2p_{\bar{\nu_e}}+p_{\gamma})_{\mu}
(p_{e^+}C_{11}-p_{\bar{\nu_e}}C_{11}
$$
$$
\hspace{0.2cm}+p_{\gamma}C_{12})^{}_{\nu}+(p_{e^+}C_{11}-p_{\bar{\nu_e}}C_{11}+p_{\gamma}
C_{12})_{\mu}(2p_{e^+}-2p_{\bar{\nu_e}}+p_{\gamma})_{\nu}
$$
$$
\hspace{0.2cm}-(p_{e^+}C_{11}-p_{\bar{\nu_e}}C_{11}+p_{\gamma}
C_{12})^{\rho}(2p_{e^+\rho}
g_{\mu\nu}-2p_{\bar{\nu_e}\rho}g_{\mu\nu}
$$
$$
\hspace{0.2cm}+p_{\gamma\rho}g_{\mu\nu}+i\varepsilon_{\mu\rho\nu\sigma}
p_{\gamma}^{\sigma})+[2(p_{e^+}-p_{\bar{\nu_e}})_{\mu}(p_{e^+}
-p_{\bar{\nu_e}})_{\nu}$$
$$
\hspace{0.2cm}-(p_{e^+}-p_{\bar{\nu_e}})^{2}g_{\mu\nu}+(p_{e^+}-
p_{\bar{\nu_e}})_{\mu}p_{\gamma\nu}-g_{\mu\nu}p_{e^+}.p_{\gamma}
$$
$$
\hspace{0.2cm}+g_{\mu\nu}p_{\bar{\nu_e}}.p_{\gamma}+p_{\gamma\mu}(p_{e^+}
-p_{\bar{\nu_e}})_{\nu}-i\varepsilon_{\mu\rho\nu\sigma}
(p_{e^+}-p_{\bar{\nu_e}})^{\rho}
$$
$$
\hspace{0.2cm}p_{\gamma}^{\sigma}]C_0\bigr\},\hspace{4.52cm}
\eqno(A1)
$$
$$                                                                   
\Gamma^{(2c)}_{\mu\nu}(\Pi)=
c_f\frac{M^{}_Wm_tm_t'}{6\pi^2F^{}_{\Pi}}\sqrt{2\sqrt{2}\pi G^{}_F\alpha}
\bigl\{2[(p_{e^+}-p_{\bar{\nu_e}})_{\mu}\hspace{0.5cm}
$$
$$
\hspace{0.2cm}(p_{e^+}-p_{\bar{\nu_e}})_{\nu}C^*_{21}+p_{\gamma\mu}
p_{\gamma\nu}C^*_{22}+(p_{e^+}-p_{\bar{\nu_e}})_{\mu}p_{\gamma\nu}C^*_{23}
$$
$$
\hspace{0.2cm}+p_{\gamma\mu}(p_{e^+}-p_{\bar{\nu_e}})_{\nu}C^*_{23}
+g_{\mu\nu}C^*_{24}]+(p_{e^+}C^*_{11}-p_{\bar{\nu_e}}C^*_{11}
$$
$$
\hspace{0.2cm}+p_{\gamma}C^*_{12})_{\mu}(2p_{e^+}-2p_{\bar{\nu_e}}
+p_{\gamma})_{\nu}
-p_{\gamma\mu}(p_{e^+}C^*_{11}-p_{\bar{\nu_e}}C^*_{11}
$$
$$
\hspace{0.2cm}+p_{\gamma}C^*_{12})_{\nu}+g_{\mu\nu}(p_{e^+}.p_{\gamma}C^*_{11}
-p_{\bar{\nu_e}}.p_{\gamma}C^*_{11})
$$
$$
\hspace{0.4cm}-i\varepsilon_{\mu\rho\sigma\nu}[(p_{e^+}-p_{\bar{\nu_e}})
C^*_{11}+p_{\gamma}C^*_{12}]^{\rho}p_{\gamma}^{\sigma}\bigr\},
\eqno(A2)
$$
$$                                                                  
\Gamma^{(2d)}_{\mu\nu}=c_f\frac{M_Wm_tm_t'}{4\pi^2F_\Pi}\sqrt{2\sqrt{2}\pi 
G_F\alpha}\hspace{1cm}\hspace{2.4cm}
$$
$$
\hspace{0.2cm}\frac{B_1(p_t+p_{\bar{b}},m_t,m_b)+B_0(p_t+p_{\bar{b}},m_t,m_b)}
{M^2_W}\hspace{1cm}
$$
$$
\hspace{0.2cm}\bigl\{(p_{e^+}-p_{\bar{\nu_e}})_{\mu}(p_{e^+}
-p_{\bar{\nu_e}})_{\nu}-p_{\gamma\mu}p_{\gamma\nu}\hspace{2cm}
$$
$$
\hspace{0.2cm}-g_{\mu\nu}(p_{e^+}
-p_{\bar{\nu_e}})^2\bigr\}\hspace{4.6cm}
\eqno(A3)
$$
$$                                                                  
-i\Sigma_{\mu}(\Pi)=
c_f\frac{M_Wm_tm_t'}{8\pi^2F_{\Pi}}\sqrt{2\sqrt{2}G_F}
(p_t+p_{\bar{b}})_{\mu}\hspace{2.4cm}
$$
$$
\hspace{0.4cm}\bigl\{B_1(p_t+p_{\bar{b}},m_t,m_b)+B_0(p_t+p_{\bar{b}},m_t,m_b)
\bigr\}
\hspace{2cm}
\eqno(A4)
$$
$$
C_{ij}=C_{ij}(p_{\bar{\nu_e}}-p_{e^+},-p_{\gamma},m_t,
m_b,m_b)~~~~
$$
$$
C^*_{ij}=C_{ij}(p_{e^+}-p_{\bar{\nu_e}},p_{\gamma},
m_b,m_t,m_t)\,,
\eqno(A5)
$$
where $C_{ij}$'s are the standard 3-point functions given in
Ref.\cite{PV}. 

The expressions for $\Gamma^{(2b)}_{\mu\nu}(\Pi_t)$, 
$\Gamma^{(2c)}_{\mu\nu}(\Pi_t)$, $\Gamma^{(2d)}_{\mu\nu}(\Pi_t)$, 
and $\sigma_{\mu}(\Pi_t)$ can be obtained by simply replacing $m'_t$ by 
$m_t-m'_t$, $F_{\Pi}$ by $F_{\Pi_t}$ and taking $c_f=1$.  

\null

\onecolumn
\newpage

\section*{TABLES}
\null\noindent
~~{\bf Tbale 1:} The masses $m_{\Pi_t}$, $m_{\Pi}$  
and the decay constants $F_{\Pi_t}$, $F_{\Pi}$ of the top-pion and the 
technipion in the TC2-I and TC2-II models.
\vspace{0.1in}
\begin{center}
\doublerulesep 0.8pt
\tabcolsep 0.1in
\begin{tabular}{||c|c|c|c|c||}\hline\hline
Model &  \multicolumn{2}{|c|}{top-pion~$\Pi_t$} 
& \multicolumn{2}{|c|}{technipion~$\Pi$}\\
\cline{2-5}
 & $m_{\Pi_t}$ (GeV) & $F_{\Pi_t}$ (GeV) & $m_{\Pi}$ (GeV) & 
$F_{\Pi}$ (GeV)\\
\hline
TC2-I & $150\sim 380$ & 50 & $100\sim 220$ & 123\\
\hline
TC2-II & $150\sim 380$ & 50 & $100\sim 220$ & 40\\
\hline\hline
\end{tabular}
\end {center}

\vspace{0.5cm}
 
{\bf Table 2:} Top-pion corrections to the
$e^+\gamma \rightarrow t\bar{b}\nu_e$ production cross section $\Delta
\sigma_{\Pi_t}$ and the total production cross section $\sigma=\sigma_{0}
+\Delta \sigma_{\Pi_t}+\Delta \sigma_{\Pi}$ in model TC2-I with
$m_{\Pi}=220$ GeV and various values of $m_{\Pi_t}$. The technipion
corrections are negligibly small. The tree level production cross section
$\sigma_0=3.25 fb$ for $\sqrt{s}=0.5 {\rm TeV}, \sigma_0=14.73 fb$ for $
\sqrt{s}=1.6 {\rm TeV}$.
\vspace{0.1in}
\begin{center}
\doublerulesep 0.8pt
\tabcolsep 0.05in
\begin{tabular}{||c|c|c|c|c|c|c||}\hline \hline
$\sqrt{s}$ & \multicolumn{3}{|c|}{$m'_t=5 GeV$} & \multicolumn{3}{|c|}
{$m'_t=20 GeV$}\\
\cline{2-7}
$(TeV)$ & $m_{\Pi_t} (GeV)$ & $\Delta\sigma_{\Pi_t} (fb)$ & $\sigma (fb)$
 & $m_{\Pi_t} (GeV)$ & $\Delta\sigma_{\Pi_t} (fb)$ & $\sigma(fb)$ \\
 \hline
    &150 &0.0025 &3.25 &150 &-0.004 &3.25 \\
    &180 &0.083  &3.26 &180 &0.06  &3.26\\
0.5 &200 &0.081   &3.33 &200 &0.062  &3.31\\
    &240 &0.12   &3.37 &240 &0.096  &3.35\\
    &300 &0.098  &3.35 &300 &0.083  &3.33\\
    &380 &0.058  &3.31 &380 &0.052  &3.30\\
\hline
    &150 &-0.12  &14.61 &150 &-0.11 &14.62 \\
    &180 &-0.11  &14.62 &180 &-0.099 &14.63\\
1.6 &200 &0.05   &14.78 &200 &0.039 &14.77\\
    &240 &0.086   &14.82 &240 &0.061 &14.79\\
    &300 &0.082   &14.81 &300 &0.058 &14.79\\
    &380 &0.05   &14.78 &380 &0.042 &14.77\\
\hline\hline
\end{tabular}
\end{center}
\newpage
{\bf Table 3:} Top-pion and technipion corrections to the
$e^+\gamma \rightarrow t\bar{b}\nu_e$ production cross section
$\Delta \sigma_{\Pi_t},\Delta \sigma_{\Pi}$ and
the total production cross section $\sigma=\sigma_{0}
+\Delta \sigma_{\Pi_t}+\Delta \sigma_{\Pi}$
in model TC2-II with $\sqrt{s}=0.5,1.6$ TeV and
various values of $m_{\Pi_t},m_{\Pi}, m'_t$.
The tree level production cross section
$\sigma_0=3.25 fb$ for $\sqrt{s}=0.5 {\rm TeV}, \sigma_0=14.73 fb$ for $
\sqrt{s}=1.6 {\rm TeV}$.
\vspace{0.1in}
\begin{center}
\doublerulesep 0.8pt
\tabcolsep 0.05in
\begin{tabular}{|c|c|c|c|c|c|c|c|c|}\hline \hline
$\sqrt{s}$ &$m_{\Pi}$ &$m_{\Pi_t}$ &
\multicolumn{3}{|c|}{$m'_t=5 GeV$} & \multicolumn{3}{|c|}
{$m'_t=20 GeV$}\\
\cline{4-9}
$(TeV)$ & $(GeV)$&$(GeV)$ & $\Delta\sigma_{\Pi_t} (fb)$ &
$\Delta\sigma_{\Pi} (fb)$& $\sigma (fb)$
& $\Delta\sigma_{\Pi_t} (fb)$ & $\Delta\sigma_{\Pi} (fb)$&
$\sigma(fb)$ \\
 \hline
    &    &150 &0.0025 &         &3.25  &-0.004  &          &3.25 \\
    &    &180 &0.0083 &         &3.26  &0.006   &          &3.26\\
    &100 &200 &0.081  &-0.0007  &3.33  &0.062   &-0.0008   &3.31\\
    &    &240 &0.12   &         &3.37  &0.096   &          &3.35\\
    &    &300 &0.098  &         &3.35  &0.083   &          &3.33\\
    &    &380 &0.058  &         &3.31  &0.052   &          &3.30\\
 \cline{2-9}
0.5 &    &150 &0.0025 &         &3.77  &-0.004  &          &6.63 \\
    &    &180 &0.0083 &         &3.78  &0.006   &          &6.64\\
    &220 &200 &0.081  &0.52     &3.85  &0.062   &3.38      &6.71\\
    &    &240 &0.12   &         &3.89  &0.096   &          &6.75\\
    &    &300 &0.098  &         &3.87  &0.083   &          &6.73\\
    &    &380 &0.058  &         &3.83  &0.052   &          &6.69\\
\hline
    &    &150 &-0.12 &         &14.56  &-0.11   &          &14.46 \\
    &    &180 &-0.11 &         &14.57  &-0.099  &          &14.47\\
    &100 &200 &0.05  &-0.047   &14.73  &0.039   &-0.16     &14.61\\
    &    &240 &0.086 &         &14.77  &0.061   &          &14.63\\
    &    &300 &0.082 &         &14.76  &0.058   &          &14.63\\
    &    &380 &0.05  &         &14.73  &0.042   &          &14.61\\
 \cline{2-9}
1.6 &    &150 &-0.12 &         &16.39  &-0.11   &          &26.14 \\
    &    &180 &-0.11 &         &16.40  &-0.099  &          &26.15\\
    &220 &200 &0.05  &1.78     &16.56  &0.039   &11.53     &26.31\\
    &    &240 &0.086 &         &16.60  &0.061   &          &26.35\\
    &    &300 &0.082 &         &16.59  &0.058   &          &26.34\\
    &    &380 &0.05  &         &16.56  &0.042   &          &26.31\\
\hline\hline
\end{tabular}
\end{center}

\vspace{5cm}

\newpage 
\begin{figure}[h]
\vspace*{13cm}
\includegraphics{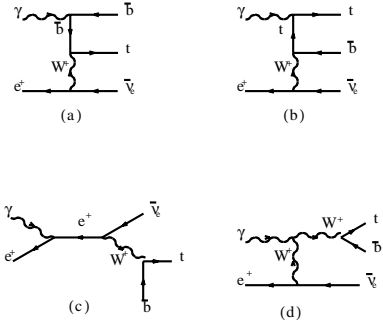}
\vspace{-6.0cm}
\caption[]{
Tree-level Feynman diagrams contributing from various technicolor models
to the process $e^{+} \gamma \rightarrow t \overline{b} \nu_e$
The dashed lines denote color-singlet technipions or top-pions. 
}
\end{figure}

\vspace{0.5cm}
\begin{figure}[h]
\vspace*{13cm}
\includegraphics{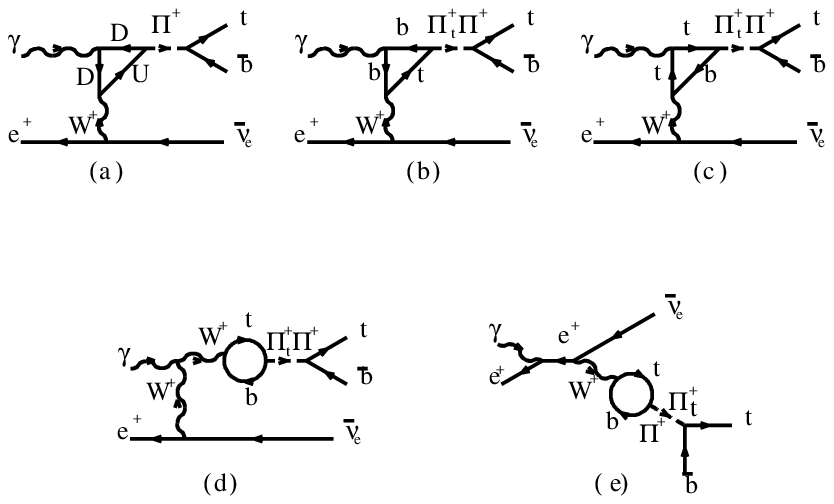}
\vspace{-6.0cm}
\caption[]{
One-loop Feynman diagrams for resonance contributing from various 
technicolor models to the process $e^{+} \gamma \rightarrow t \overline{b} 
\nu_e$ The dashed lines denote color-singlet technipions or top-pions. 
}
\end{figure}

\newpage
\begin{figure}[h]
\vspace*{15cm}
\includegraphics{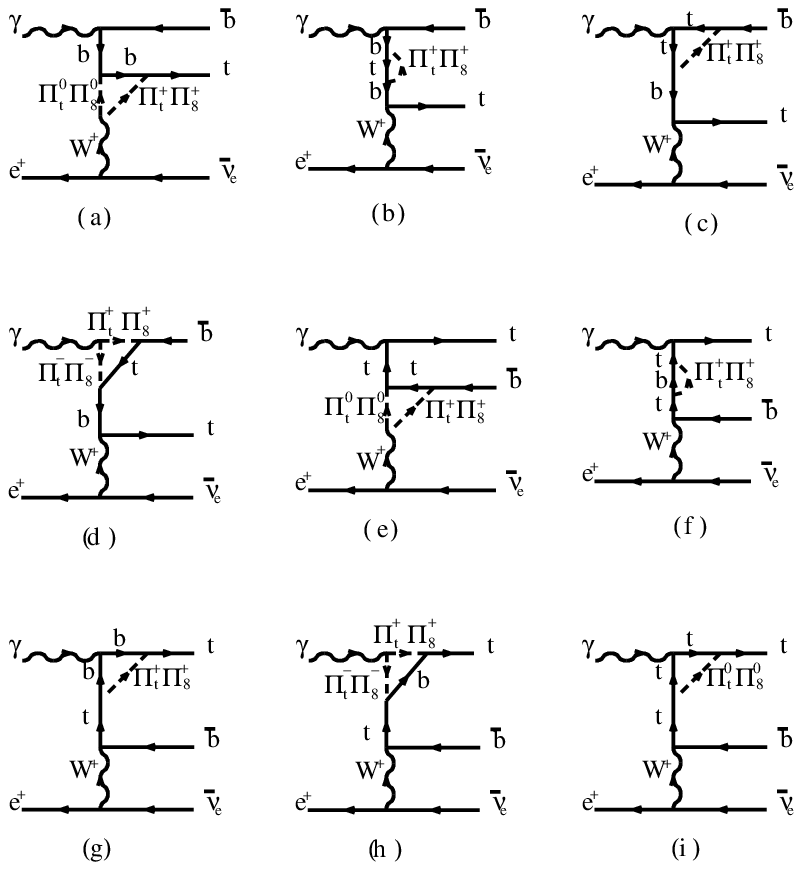}
\vspace{-4.0cm}
\caption[]{
One-loop Feynman diagrams for non-resonance contributing from various 
technicolor models to the process $e^{+} \gamma \rightarrow t \overline{b} 
\nu_e$ The dashed lines denote color-singlet technipions or top-pions. 
}
\end{figure}

\newpage
\begin{figure}[h]
\vspace*{10cm}
\includegraphics{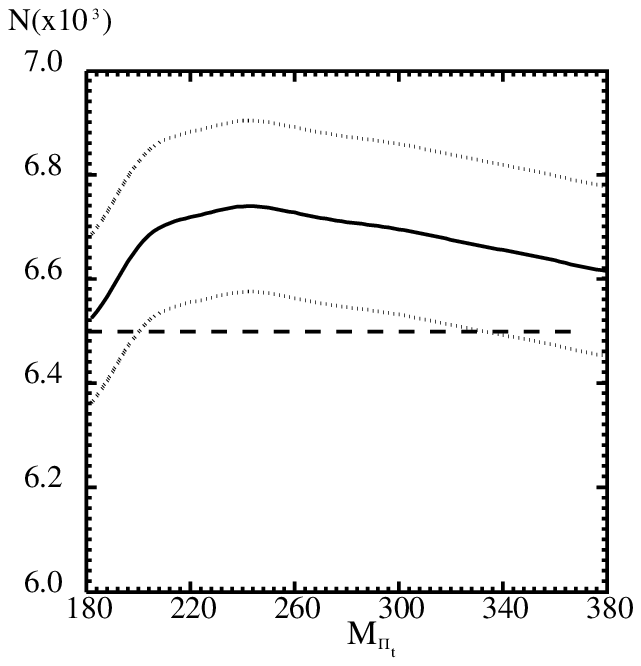}
\vspace{-1.0cm}
\caption[]{
The number of events $N$ (4-year run at the TESLA) versus $M_{\Pi_t}$ 
for $m'_t=5$ GeV and $M_{\Pi}=220$ GeV at $\sqrt{s}=0.5$ TeV
in model TC2-I. The solid line is the total number of events, the dashed line 
is the number of events corresponding to the tree-level standard model
contribution, the dotted lines indicate the statistical uncertainty bounds 
at $95\%$ C.L.}
\end{figure}

\begin{figure}[h]
\vspace*{10cm}
\includegraphics{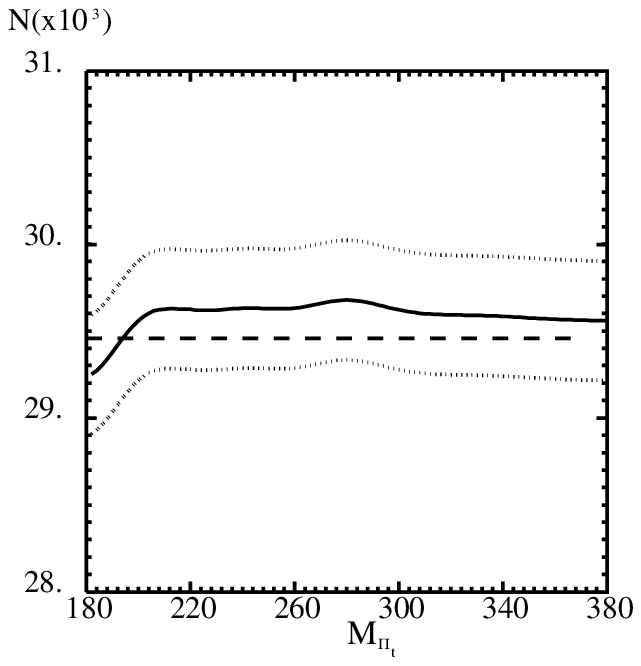}
\vspace{-1.0cm}
\caption[]{
The number of events $N$ (4-year run at the TESLA) versus $M_{\Pi_t}$ 
for $m'_t=5$ GeV and $M_{\Pi}=220$ GeV at $\sqrt{s}=1.6$ TeV
in model TC2-I. The solid line is the total number of events, the dashed line 
is the number of events corresponding to the tree-level standard model
contribution, the dotted lines indicate the statistical uncertainty bounds 
at $95\%$ C.L.}
\end{figure}

\begin{figure}[h]
\vspace*{10cm}
\includegraphics{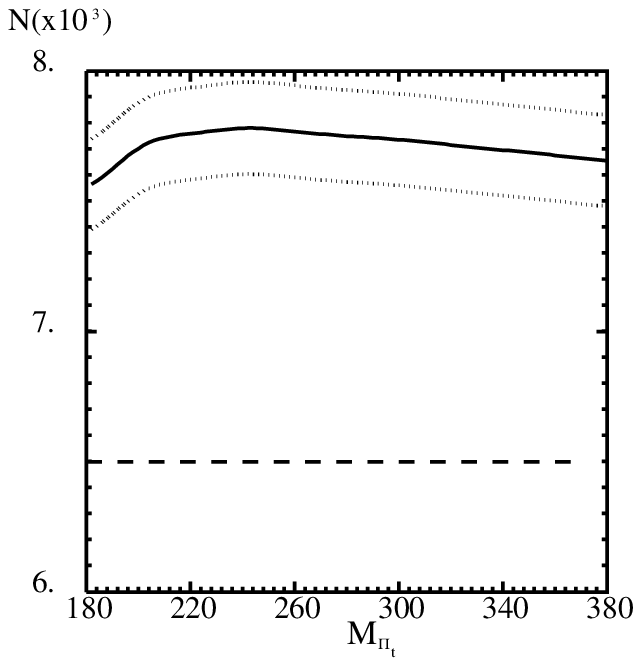}
\vspace{-1.0cm}
\caption[]{
The number of events $N$ (4-year run at the TESLA) versus $M_{\Pi_t}$ 
for $m'_t=5$ GeV and $M_{\Pi}=220$ GeV at $\sqrt{s}=0.5$ TeV
in model TC2-II. The solid line is the total number of events, the dashed line 
is the number of events corresponding to the tree-level standard model
contribution, the dotted lines indicate the statistical uncertainty bounds 
at $95\%$ C.L.}
\end{figure}

\begin{figure}[h]
\vspace*{10cm}
\includegraphics{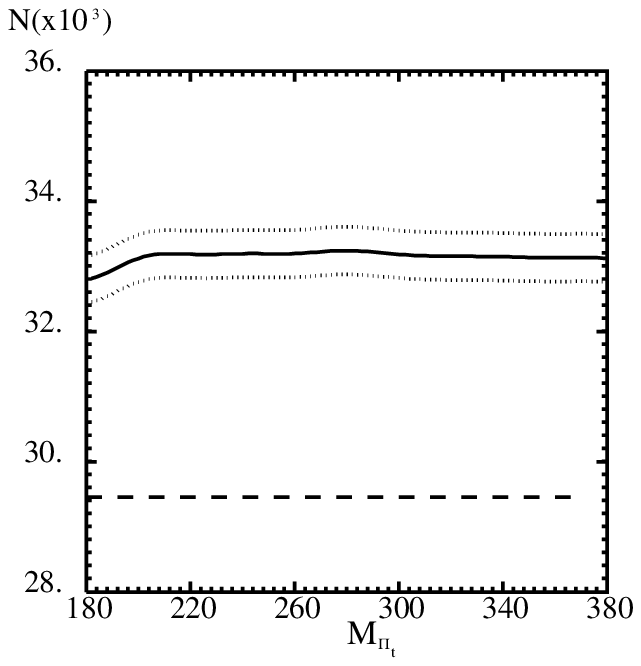}
\vspace{-1.0cm}
\caption[]{
The number of events $N$ (4-year run at the TESLA) versus $M_{\Pi_t}$ 
for $m'_t=5$ GeV and $M_{\Pi}=220$ GeV at $\sqrt{s}=1.6$ TeV
in model TC2-II. The solid line is the total number of events, the dashed line 
is the number of events corresponding to the tree-level standard model
contribution, the dotted lines indicate the statistical uncertainty bounds 
at $95\%$ C.L.}
\end{figure}

\end{document}